\documentclass[english,twocolumn, prl, noeprint, aps, superscriptaddress, numerical, footinbib]{revtex4-1}

\usepackage{geometry}
\usepackage{amsmath}
\usepackage{babel}
\usepackage{graphicx}
\usepackage{microtype}
\usepackage{hyperref}
\hypersetup{
  colorlinks = true,
  linkcolor = blue,
  urlcolor = blue,
  citecolor = blue
}

\usepackage{xcolor}
\usepackage{xspace}

\newcommand{\om}{\omega_0}
\newcommand{\piLS}{phononic Lamb shift\xspace}
\newcommand{\fexc}{\eta}

\newcommand{\detun}{\Delta}

\newcommand{\rhoLiBEC}{\rho_\mathrm{2D}}
\newcommand{\coup}{\alpha_\mathrm{IB}}
\newcommand{\dSelf}{\delta_\mathrm{self}}
\newcommand{\VB}{V_\mathrm{B}}
\newcommand{\VI}{V_\mathrm{I}}
\newcommand{\dSelfF}{\delta_\mathrm{self}^\mathrm{F}}
\newcommand{\dSelfB}{\delta_\mathrm{self}^\mathrm{B}}
\newcommand{\aIB}{a_\text{IB}}

\DeclareMathOperator\erf{erf}

\newcommand{\nm}{~\mathrm{nm}}
\newcommand{\ms}{~\mathrm{ms}}
\newcommand{\Hz}{~\mathrm{Hz}}
\newcommand{\um}{~\mu\mathrm{m}}
\newcommand{\bohr}{~a_0}
\newcommand{\THz}{~\mathrm{THz}}
\newcommand{\kHz}{~\mathrm{kHz}}

\def\0{^{\vphantom{\dagger}}}

\begin{document}


\begin{abstract}
  The quantum vacuum fundamentally alters the properties of embedded particles. In contrast to classical empty space, it allows for creation and annihilation of excitations. For trapped particles this leads to a change in the energy spectrum, known as Lamb shift. 
Here, we engineer a synthetic vacuum building on the unique properties of ultracold atomic gas mixtures. This system makes it possible to combine high-precision spectroscopy with the ability of switching between empty space and quantum vacuum.
We observe the phononic Lamb shift, an intruiguing many-body effect orginally conjectured in the context of solid state physics. Our study therefore opens up new avenues for high-precision benchmarking of non-trivial theoretical predictions in the realm of the quantum vacuum.
\end{abstract}

\title{Observation of the phononic Lamb shift with a synthetic vacuum}

\author{T.~Rentrop}
\email{LambShift@matterwave.de}
\author{A.~Trautmann}
\author{F.~A.~Olivares}
\author{F.~Jendrzejewski}
\affiliation{Kirchhoff-Institut f\"ur Physik, Universit\"at Heidelberg, Im Neuenheimer Feld 227, 69120 Heidelberg, Germany.}
\author{A.~Komnik}
\affiliation{Institut f\"ur Theoretische Physik, Universit\"at Heidelberg, Philosophenweg 12,  69120 Heidelberg,  Germany.}
\author{M.~K.~Oberthaler}
\affiliation{Kirchhoff-Institut f\"ur Physik, Universit\"at Heidelberg, Im Neuenheimer Feld 227, 69120 Heidelberg, Germany.}
\maketitle 

For the electron in the hydrogen atom the coupling to virtual photons leads to the mass renormalization and the Lamb shift, which has been measured with unprecedented precision \cite{Lamb1947,Bethe1947, Berkeland1995, Niering2000} . Similar effects take place in semiconductors where electrons couple to phononic excitations \cite{Landau1933, Feynman1955}. The increase of the effective mass has been observed in these systems \cite{Hodby1976}. A quantitative measurement of the predicted phononic Lamb shift \cite{Platzman1961, Mitra1987}, which is defined for an electron bound to a donor ion or some other attractive center, is still missing. The main reason is  that such a system is notoriously difficult to engineer in solid state materials, due to uncontrolled disorder effects \cite{Simmonds1974}. Here, we realize a model system for such a phononic coupling with ultracold atoms, where we have full control over the phononic background as well as the bound state. If the interaction is sufficiently weak, this process can be quantitatively described within the Fr\"ohlich model \cite{Froehlich1954, Tempere2009, Grusdt2014b, Hofstetter2015}. In this regime, the phononic Lamb shift is minute and therefore it is an experimental challenge to resolve it accurately.

We perform the experiments in a strongly imbalanced mixture of different ultracold atomic gases. The impurities, the minority species, are well localized in a tight optical trap, in analogy to the electron bound to a donor ion. The majority species is a Bose-Einstein condensate (BEC), which plays the role of the quantum vacuum. We observe the \piLS directly by high-resolution spectroscopy of the two lowest energy levels of the bound impurity employing motional Ramsey spectroscopy \cite{Scelle2013}. Previously demonstrated methods for the investigation of the polaronic effects are quantum phase revival  \cite{Will2011}, multiband spectroscopy \cite{Heinze2011} and oscillations in a shallow traps \cite{Nascimbene2009, Catani2012}.  For the quantitative analysis we build on the ability to remove the BEC, i.e. switching off the quantum vacuum; a feature that does not exist in quantum electrodynamics (QED) experiments \cite{Brune1994, Fragner2008} and semiconductor systems.  This ability makes the energy shifts due to the quantum vacuum directly accessible. 

Our experimental platform allows for the realization of fermionic as well as bosonic impurities. The latter experience a significant enhancement of the Lamb shift via Bose amplification. This approach enables a systematic study of the \piLS in the weak coupling regime. All experimental results show quantitative agreement with the predictions of the Fr\"ohlich Hamiltonian.
 
\begin{figure*}[t!]
\includegraphics[width = 180 mm]{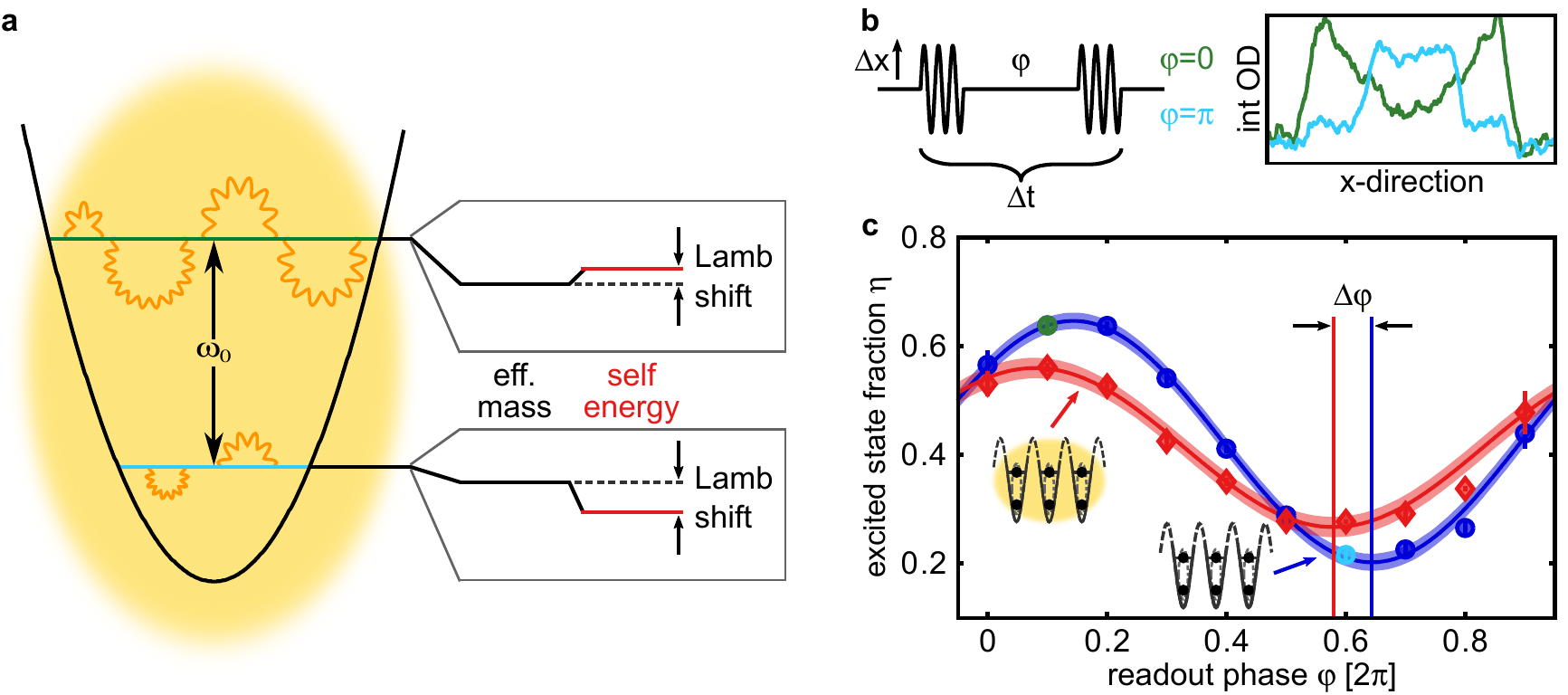}
\caption{\textbf{Detection of the self-energy shift. a} Interaction of impurities with phonons of the BEC causes energy shifts of the impurity's external states. \textbf{b} Shaking the optical lattice couples the external states and allows the preparation of a superposition of the two lowest external states. Changing the evolution time $\Delta t$ results in a Ramsey fringe in the population of the excited state, which is read out via band mapping. Typical profiles of the resulting absorption images are depicted. \textbf{c} An exemplary interference fringe of \textsuperscript{6}Li after an evolution time of  1.1 ms is shown. In case of an evolution with background the amplitude is reduced due to decoherence and the fringe is shifted, showing an increased energy difference. The shaded area corresponds to the $1\sigma$ confidence levels of the phase estimation. 
} \label{FigRam}
\end{figure*}

In our experiment, the confinement of the fermionic impurity atoms (\textsuperscript{6}Li) is realized by a deep one-dimensional (1D) species-selective optical lattice. It can be approximated by independent harmonic oscillators (see Fig.~\ref{FigRam} a). The weak confinement in the other two directions arises from a shallow optical trap generating an array of two-dimensional (2D) gases \cite{SuppNote}. After initial preparation in the 1D ground state of the harmonic oscillator we create a coherent superposition in the two lowest energy states of the oscillator by shaking the optical lattice \cite{Scelle2013}. In the subsequent time evolution the excited motional state will accumulate a phase with respect to the ground state corresponding to its energy difference. The accumulated phase difference can be mapped onto an observable population by applying an additional shaking pulse. In our system of periodically arranged harmonic oscillators, the population can be accessed by the band mapping technique, see Fig.~\ref{FigRam} b, \cite{Greiner2001}.
In the absence of the BEC we observe the corresponding Ramsey fringe by changing the phase of the second shaking pulse and extracting the population of the excited state, see Fig.~\ref{FigRam} c, blue curve. We clearly observe a fringe shift (Fig.~\ref{FigRam} c, red curve) in the presence of the BEC (\textsuperscript{23}Na), which indicates an energy shift. The sign of the phase shift $\Delta \phi$ corresponds to an increase of the energy difference, which is at odds with a naive interpretation of an increased mass (see Fig.~\ref{FigRam} a).

This qualitative deviation from the effective mass increase predicted for particles coupled to a quantum vacuum is the manifestation of the additional effect of the Lamb shift for bound particles. This is well captured by the Fr\"ohlich Hamiltonian \cite{Tempere2009, Grusdt2014b}:
\begin{equation}
 H = \sum \limits_{k} E_k \, \hat{a}^\dag_k \hat{a}_k\0 + \sum \limits_{q} \omega_q \, \hat{b}^\dag_q \hat{b}_q\0 + \sum \limits_{k,q \neq 0}V_q \hat{a}^\dag_{k+q} \hat{a}_{k}\0 (\hat{b}_q\0 + \hat{b}^\dag_{- q})\mbox{,}
\end{equation}

where $E_k$ ($\omega_q$) represent the energy levels of the uncoupled impurities (phonons). The impurity (phonon) creation and annihilation operators are denoted $\smash{ \hat{a}^\dag_k}$($\smash{ \hat{b}^\dag_q}$) and $\smash{ \hat{a}_k\0}$($\smash{ \hat{b}_q\0}$). The third term arises from the density-density interaction between the impurities and the BEC. It describes the change in momentum of the impurity atom via the absorption or emission of a phonon. We emphasize that the coupling strength $V_q$ captures the contact interaction in our system, which is different from the long-range Coulomb interaction of an electron in a semiconductor. Nevertheless, the same phenomenology  of the effective mass and  the phononic Lamb shift exists in both cases. In order to access them, we calculate the lowest-order self-energy \cite{SuppNote}. The effective mass is extracted from the self-energy of  an unbound particle \cite{SuppNote}, leading to an energy shift  (Fig.~\ref{FigRam} a). It corresponds to a slow down in the oscillations. However,  the effects of the confinement go far beyond the simple effective mass approximation. We term the deviation \piLS in close analogy with the Lamb shift in the hydrogen atom (Fig.~\ref{FigRam} a).

Generally, ground and excited state have different self-energies. The ground state experiences a larger shift since its extension is smaller, resulting in a higher density. Therefore the density-density interaction causes stronger coupling and thereby a stronger shift. Our detection method reveals this differential energy shift $\Delta \omega$ between the ground and the excited state:
\begin{equation}
\dSelf=  \frac{\Delta \omega}{\om} = \coup \cdot f(\rhoLiBEC, a, \xi).
\end{equation}
$\om$ is the energy difference of the two impurity states in empty space, $\coup$ is the coupling strength between the impurity and the background for a confined geometry \cite{SuppNote}. The confinement geometry of the impurity is taken into account by the function $f$, which depends on the 2D density of the impurities $\rhoLiBEC$, the harmonic oscillator length $a$, and the healing length of the background condensate $\xi$.

\begin{figure}[t]
\includegraphics[width = 86 mm]{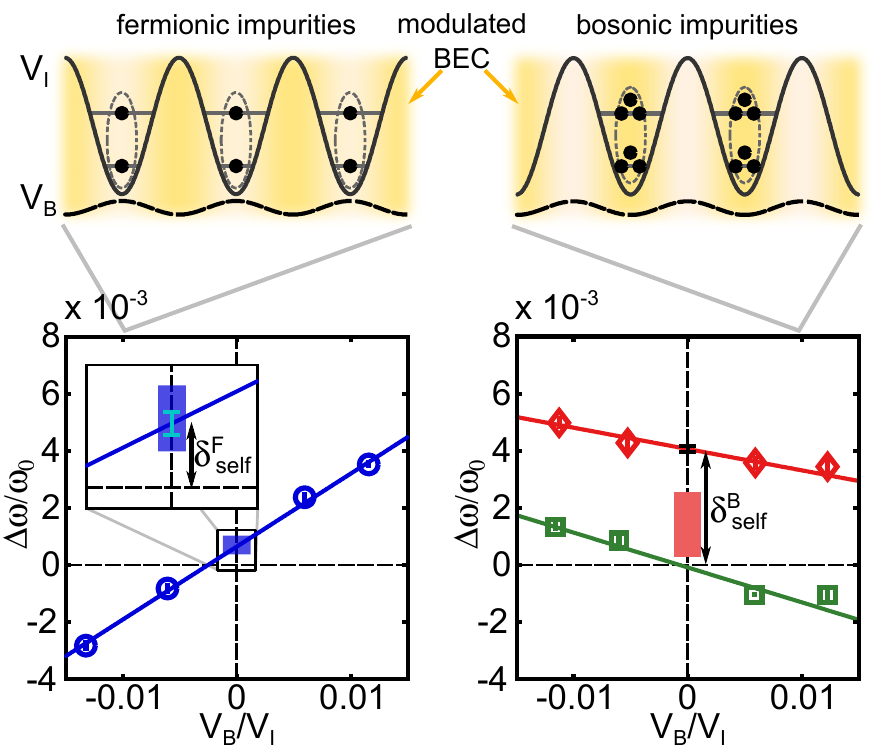}
\caption{ \textbf{Observation of the self-energy.} The residual modulation of the density of the background BEC leads to an additional energy shift of the impurity energy level. This can be identified by altering the ratio $\VB / \VI$ through a change of the detuning of the optical lattice while keeping the potential depth for the impurity constant. A clear linear dependence on $\VB/\VI$ is observed. Interpolating the measurement data to $\VB=0$ allows for extraction of the self-energy shift. The positive offset is a direct signature of the phononic Lamb shift, since it contradicts the negative energy shift expected from the increased effective mass. The shaded areas show the theoretically predicted shift in the Fr\"ohlich scenario, taking into account the uncertainties of the experimental parameters. 
We find perfect agreement without free parameters for fermionic impurites (blue). 
In case of a non-condensed bosonic impurity cloud (green squares), no shift is observable in  agreement with the theoretical expection. For a condensed bosonic impurity cloud (red diamonds), we observe a significant change of the self-energy. The theoretical expectation, not taking into account thermal excitations ($T=0$), is depicted as the red shaded area.}
\label{FigLatt}
\end{figure}

For a quantitative comparison care has to be taken since the employed near-resonance lattice for lithium induces also a weak modulation of the BEC. As shown in the upper row of Fig.~\ref{FigLatt} it modifies the effective confinement of the impurities due to mean field interactions between the two species \cite{SuppNote}. These effects are isolated by performing the Ramsey spectroscopy at different detunings of the lattice. By going closer to the lithium resonance ($\detun = 2\pi\cdot 0.3 \THz$) the potential depth $V_I$ is kept constant ($\om \approx 2\pi\cdot 27 \kHz$) by reducing the intensity accordingly. Since the transition frequency for the background is far-detuned ($\detun = 2\pi\cdot 63 \THz$), the corresponding potential $V_B$ is reduced and the background modulation suppressed. We observe that the frequency shifts have a finite offset at the interpolated limit of zero background modulation. This is exactly $\dSelf$. The fit yields for the fermionic \textsuperscript{6}Li (blue dots) the relative shift $\dSelfF= (6.4\pm1.2)\cdot10^{-4}$, where the error bar corresponds to the 68\,\% confidence interval. The shaded blue area  in Fig.~\ref{FigLatt} indicates the range of theoretically predicted values, taking into account the uncertainties in the density distribution of the clouds \cite{SuppNote}. Our observation confirms the theoretically predicted self-energy shifts. So the increased energy spacing is indeed a direct manifestation of the phononic Lamb shift in our system.

\begin{figure}[t]
\includegraphics[width = 86mm]{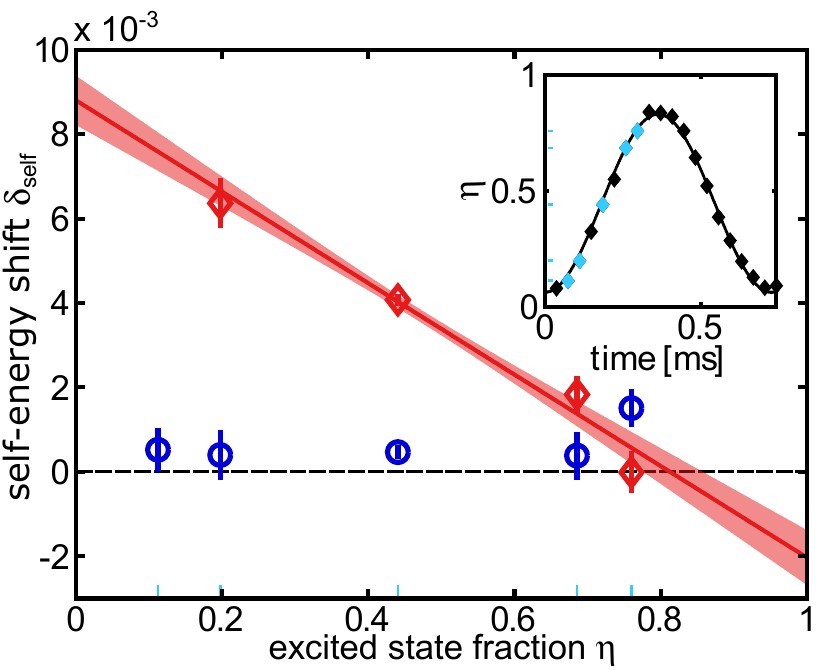}
\caption{\textbf{Population dependence of the self-energy.} The self-energy shift depends on the relative state population $\fexc$ as ground and excited state experience different energy shifts. This can be observed in a Ramsey sequence with an unequal superposition during free time evolution, keeping the other relevant parameters constant. Different $\fexc$ can be obtained by shaking the optical lattice for different times (inset). Experiments with bosonic impurities are shown by red diamonds. From these data we extract a critical $\fexc = 0.81 \pm 0.07$ in good agreement with theoretical predictions. For fermionic impurities (blue circles) we do not observe any dependence on $\fexc$ as expected from our calculations.}
\label{FigPD}
\end{figure}

Employing bosonic impurities results in a larger Lamb shift due to bosonic enhancement. We employ this feature to boost the Lamb shift in experiments with bosonic impurities (\textsuperscript{7}Li) that are condensed ($\approx 60\%$ condensate fraction).  The results (red diamonds) are displayed in Fig.~\ref{FigLatt}. We observe that  the effect of the background modulation is inverted and smaller than for fermionic impurities, as the interspecies scattering length changes sign and is reduced by a factor of three. Thus, the energy shift for a single boson is predicted to be ten times smaller than for fermionic \textsuperscript{6}Li. Nevertheless, the observed energy shift is amplified by a few thousands of bosons in the ground state leading to $\dSelfB = (4.1\pm0.1)\cdot10^{-3}$, exceeding theoretical prediction. A quantitative comparison between theory and experiment is difficult in this specific case. First, the overlap between the two species is highly sensitive to details in the trap geometries, limiting its control. Second, finite temperature effects represent a challenge to the theoretical description.
For comparison we also performed the same experiments with a non-condensed cloud of bosons (green squares), leading to smaller number of atoms per quantum state, and a weakened coupling as we operated at a smaller background density. As expected from the theoretical predictions, we do not observe any shift in this regime.

For a quantitative comparison between theory and experiment for bosonic impurities, we measure the self-energy shift as a function of different populations of ground and excited state. 
Since the self-energy shift is proportional to the occupation number for bosons there exists a specific relative population $\fexc$, where the energy shifts are equal for both levels. For our experimental set-up, this implies a vanishing phase shift for $\fexc = 0.87$. By varying the length of the first Ramsey pulse, the relative occupation of the excited state during time evolution is changed. The inset in Fig.~\ref{FigPD} shows a Rabi cycle, which we use for the calibration of $\fexc$. Fig.~\ref{FigPD} depicts the dependence of $\dSelf$ on the excited fraction and confirms the prediction.  While the experiments with fermions (blue) do not reveal any significant change, the data with bosonic impurities (red) show a clear linear dependence with a crossing point at the expected value. We emphasize that this crossing point depends only weakly on the overlap between the impurities and the BEC. As such it puts the theoretical predictions to a precise test. In contrast, the experiments with fermions (blue) do not reveal any significant change in agreement with our theoretical treatment.

Our results are in the realm of implementation of analog quantum simulators on the basis of ultracold gas mixtures \cite{Bloch2008} . We open up an avenue for benchmarking a new class of many-body theories, comprising not only different mutually interacting particles, but also interactions of their collective excitations. Furthermore, it captures the most important non-relativistic features of the QED. In this way we realize a new experimental platform for quantitative exploration of such fascinating phenomena as Casimir effect in non-trivial geometries as well as in dynamical situations \cite{Milonni1994BOOK, Klein2005, Wilson2011}. In a many-particle limit the background mediated interaction between the impurities can lead to a generation and detection of non-local entanglement \cite{Ramos2014}.  

\begin{acknowledgments}
 We would like to thank F.~Grusdt, E.~Demler, and R.~Schmidt for discussions. This work was supported by the Heidelberg Center for Quantum Dynamics, the ExtreMe Matter Institute and the European Commission FET-Proactive grant AQuS (Project No. 640800). F.~O. acknowledges support by the IMPRS-QD. A.~K. is supported by the Heisenberg Programme of the Deutsche Forschungsgemeinschaft (Germany) under Grant No. KO 2235/5-1. 
\end{acknowledgments}

%

\clearpage
\newgeometry{a4paper,
left=0.75in,right=0.483in,top=0.75in,bottom=1.55in
}

\onecolumngrid
~ ~
\begin{center} 
\begin{Large} \textbf{Supplementary Material} \end{Large}
\end{center}
\medskip
\smallskip
\twocolumngrid

\setcounter{section}{0}
\setcounter{subsection}{0}
\setcounter{figure}{0}
\setcounter{equation}{0}
\setcounter{NAT@ctr}{0}

\renewcommand{\figurename}[1]{FIG. }
\renewcommand{\theequation}{S\arabic{equation}}

\makeatletter
\renewcommand{\bibnumfmt}[1]{[S#1]}
\renewcommand{\citenumfont}[1]{S#1}
\renewcommand{\thefigure}{S\@arabic\c@figure} 
\makeatother
\renewcommand{\theequation}{S\arabic{equation}} 
\renewcommand\thetable{S\arabic{table}}
\renewcommand{\vec}[1]{{\boldsymbol{#1}}}

\section*{Experimental design}

In our experiment, a BEC of about 10\textsuperscript{6} sodium atoms and few 10\textsuperscript{3} to several 10\textsuperscript{4} lithium atoms are both trapped by the same two beam optical dipole trap (ODT) at 1064 nm. In all the experiments the gases are spin-polarized in the absolute hyperfine ground state. The mean trapping frequency for $^{23}$Na is $\bar{\omega}=2\pi\cdot 150 \Hz$. The temperature of the sample is $\approx$ 350 nK. We can choose to work with fermionic \textsuperscript{6}Li as well as bosonic \textsuperscript{7}Li at mean trapping frequencies of $\bar{\omega} = 2\pi\cdot 340 \Hz$ for \textsuperscript{6}Li and $\bar{\omega}=2\pi\cdot 310 \Hz$ for \textsuperscript{7}Li.   An additional optical standing wave close to optical transition for lithium ($\lambda_\mathrm{res, Li} \approx 671 \nm$) imposes a very strong confinement in one direction for lithium only. The standing wave allows the implementation of motional Ramsey spectroscopy of external impurity energy levels with a high precision \cite{S_Scelle2013}.  Due to the depth of the potential the minima can be treated as independent harmonic oscillators, resulting in multiple realizations of the experiment in a single experimental cycle. The geometry corresponds to 2D gases. For fermions the transverse extension is 10 times larger  than the longitudinal one. The background induced effect is isolated by alternately measuring the motional energy difference with and without BEC background and is detected as a phase alteration of the Ramsey readout fringes ($\Delta \phi$). The frequency change is given by $\Delta \omega = \Delta \phi / \Delta t$, where $\Delta t$ is the total evolution time, including the time of the state coupling (pulses). We emphasize that the impurity-impurity interaction is negligible, as fermions at low temperatures do not scatter and the intra-species scattering length for the bosons is 7$\bohr$ \cite{S_Salomon2002}, with $\bohr$ being the Bohr radius.

\subsection*{Species-selective optical lattice}
The species-selective optical lattice is close to D-line transitions for lithium (670 to 672 nm while the depth for Li is kept constant). It consists of two intersecting laser beams, leading to a periodicity of $d_\mathrm{lat} = 1.65 \um$ and a typical depth of 24.5 lattice recoil (E$_{\mathrm{rec}}$) for \textsuperscript{6}Li and 33\,E$_{\mathrm{rec}}$ for \textsuperscript{7}Li, corresponding to a frequency of 27.2~kHz.  

\subsection*{Coupling of motional states}
All presented measurements are performed using a Ramsey pulse scheme in the two lowest external states of the species-selective optical lattice. For coherent coupling of these states the lattice position is periodically modulated. The resulting Rabi frequency is 1.4~kHz. The finite excitation efficiency is caused by residual atoms in the excited state at the beginning of the experiment and by a weak coupling of the second excited state which is suppressed by the anharmonicity of the potential. By coupling the states for a certain time, an equal superposition of ground and excited state is created ($\pi/2$~pulse). A second $\pi/2$ pulse is used for the readout. The second pulse is shifted in time to record the phase of the oscillation. In order to measure the background induced effect on the frequency, the sodium BEC is removed by a resonant light pulse before the Ramsey sequence starts. This light pulse does not cause any observable heating of the Li sample.

\subsection*{Motional state detection}
The population detection of the external states is done by a band mapping technique \cite{S_Greiner2001}. While adiabatically reducing the lattice depth in 2~ms with a time constant $ \tau= 0.8~\mathrm{ms}$ the longitudinal confinement is turned off, allowing the atoms to expand along the lattice direction. After 10~ms of evolution an absorption image is taken. With this technique a high optical density can be obtained even for low atom numbers, corresponding to $\approx 200$ times more background than impurity atoms.

\section*{Data analysis}

\subsection*{Data taking strategy}
Each Ramsey fringe is sampled at 10 points. One scan consists of at least three runs. For each run the scans with and without background are separately fitted with a sine function and the final value for the phase shift is calculated as a weighted mean of the repetitions. Shots with extreme atom numbers, deviating lattice intensity or at wrong wavelength are discarded. If several scans can be combined, their result is calculated as the weighted mean of the individual values.

\subsection*{Background modulation}
In the Thomas-Fermi approximation the modulation of the BEC density is: $n_\mathrm{mod}=(\mu_\mathrm{B}-V_\mathrm{B})/g_\mathrm{BB}$ with $g_{\mathrm{BB}} = 4\pi\hbar^2 a_{\mathrm{BB}}/m_{\mathrm{B}}$ with $a_\mathrm{BB} = 54.54\bohr$ \cite{S_Knoop2011}. This adds to the trap potential for the impurities: $V_\mathrm{Imod} = g_\mathrm{IB} n_\mathrm{mod}$ with $g_{\mathrm{IB}} = 2\pi\hbar^2 a_{\mathrm{IB}}/m_{\mathrm{r}}$, where $a_{\mathrm{IB}} = -75\bohr$ for fermions and $a_{\mathrm{IB}} = 21\bohr$ for bosons \cite{S_Schuster2012}. The resulting impurity trapping potential is up to the constant offset: $V_\mathrm{Ieff} = V_\mathrm{I} - g_\mathrm{IB}  V_\mathrm{B}/ g_\mathrm{BB}$. In harmonic approximation $\omega \propto \sqrt{V}$, therefore $\omega_\mathrm{eff} = \om \sqrt{1 - g_\mathrm{IB}/g_\mathrm{BB}\cdot V_\mathrm{B}/V_\mathrm{I}}$. The relative shift due to the lattice is then:
\begin{equation}
\delta_\mathrm{latt}=\frac{\omega_\mathrm{eff}-\om}{\om} = -\frac{1}{2}\frac{g_{\mathrm{IB}}}{g_{\mathrm{BB}}}\frac{V_{\mathrm{B}}}{V_{\mathrm{I}}}.
\end{equation}
As $g_{\mathrm{IB}}$ is three times larger for fermionic lithium than for the bosonic isotope, this effect is more pronounced in the fermionic case. Furthermore $g_{\mathrm{IB}}$ has opposite sign for fermionic compared to bosonic lithium, inverting the effect when changing the isotope. 
This behaviour can be seen in Fig.~2. The difference of the slope for the two bosonic scenarios arises due to  a change of the effective background density $\bar{n}$. Specifically, for the systematic studies shown in Fig.~3 we keep the optical potential fixed and subtract the additional frequency shift due to background modulation.

\subsection*{Calculation of density distribution}
The density distributions are obtained numerically by taking into account the finite temperature as well as the external potentials such as the species-selective optical lattice and the harmonic trapping due to the ODT. For lithium it is necessary to include the additional potential due to the interaction with sodium. In order to capture properly the lithium occupation number in the individual sites we slice the ODT density according to the periodicity of the lattice. These occupation numbers enter the detailed calculation of the density distribution in the total confinement potential, where a 2D description is applied.

\subsection*{Experimental signal}
The band mapping technique yields the spatially averaged phase shift $\Delta \phi$. However, different parts of the lithium cloud are embedded in varying sodium densities, which leads to locally varying phase shifts $\phi({\bf r}) = \om \cdot \delta_\mathrm{self}({\bf r}) \cdot \Delta t$. These local phase shifts can be calculated within the local density approximation from Eq.~(2) in the main text. 

In the bosonic case, the observed total phase shift can then be calculated as a lithium density weighted mean:
\begin{equation}
\Delta \phi_B = \mbox{asin}\left(\frac{\int \mathrm d^3{\bf r}~n_\mathrm{Li, BEC}({\bf r}) \sin(\phi({\bf r})) }{\int \mathrm d^3{\bf r}~n_\mathrm{Li}({\bf r})}\right)
\end{equation}
For bosonic impurities only the condensed part is assumed to experience an energy shift. However, the total signal is given as a weighted sum over shifted and unshifted phase patterns. The weak interaction between the BEC and the bosonic impurites leads to long coherence times ($\tau>50\ms$), which are much longer than the Ramsey sequence, such that decoherence can be neglected in this case.

For fermionic impurities on the other hand, the interaction with the BEC is sufficiently strong that decoherence effects in the center of the cloud affect the observed signal. They can be taken into account via a position dependent relaxatio†n rate $\Gamma({\bf r})$, which was studied in detail in \cite{S_Scelle2013}. The total signal of the fermions is then given by:
\begin{equation}
\Delta \phi_F = \mbox{asin}\left(\frac{\int \mathrm d^3{\bf r}~n_\mathrm{Li}({\bf r}) e^{-\Gamma({\bf r}) t}  \sin(\phi({\bf r})) }{\int \mathrm d^3{\bf r}~n_\mathrm{Li}({\bf r})e^{-\Gamma({\bf r}) t}}\right)
\end{equation}

\section*{Theoretical treatment}
We consider two different geometries. In order to obtain the effective mass $m_\mathrm{I}^*$ we work with an unconfined (free) impurity coupled to a structureless BEC as described by Eq.~(1) in the main text. Computation of $m_\mathrm{I}^*$ can be very conveniently performed following the lines of ~\cite{S_Mahan1990BOOK}, but with different matrix elements 
\begin{eqnarray}      \label{Vk}
V_{\bf q} = \lambda [(\xi q)^2/ ((\xi q)^2 + 2)]^{1/4} \, ,
\end{eqnarray}
where $\lambda = g_{\mathrm{IB}} \, \sqrt{n_{\mathrm{BEC}}}$, see e.\,g.~\cite{S_Tempere2009, S_Grusdt2014, S_Grusdt2014b, S_Shashi2014}. 
For weak interactions one obtains $m_\mathrm{I}^*/m_\mathrm{I} = 1 + \nu \alpha + \dots$, where $\nu\approx 0.364$ and $0.336$ for ${}^{6}$Li and ${}^{7}$Li setups, respectively, and where $\alpha = \aIB^2/(\xi a_\mathrm{BB})$ is the dimensionless interaction strength.

In the second geometry we consider the impurities as being confined in a parabolic potential (energy parameter $\hbar \om$ and length parameter $a=\sqrt{\hbar/m_\mathrm{I} \om}$) in $x$-direction, and being free in all other spatial dimensions. Their eigenstates then have energies $E_{n, k} = \hbar \om (n+1/2) + \hbar^2 k^2/2m_\mathrm{I} - \mu$, where ${\bf k}$ is a 2D wave vector and $n$ denotes the respective subband of the confinement potential. The unperturbed fermion/boson Hamiltonian is then 
\begin{equation}
 H_\mathrm{I} = \sum_{n} \int \frac{\mathrm d^2 {\bf k}}{(2 \pi)^2} E_{n, k} \, \hat{a}^\dag_{n, {\bf k}} \hat{a}_{n, {\bf k}}\0 \, ,
\end{equation}
where $\hat{a}^\dag_{n, {\bf k}}$ stands for the impurity creation operator.  As $a_{\mathrm{IB}} \neq 0$ the impurities are scattered on the harmonic modes of the BEC -- the phonons. The corresponding interaction terms have been derived in the weak depletion limit in \cite{S_Tempere2009}. Adapting it to the present setup leads to the following interaction term:
\begin{eqnarray}
  H_{\mathrm{int}} &=&  \int \frac{\mathrm d^2 {\bf k}}{(2 \pi)^2} \int \frac{\mathrm d^3 {\bf q}}{(2 \pi)^3} \sum_{n_1, n_2} V_{\bf q} \, A(n_1,n_2,q_x) \, 
   \nonumber \\ 
              &&\times 
  \hat{a}^\dag_{n_1, \bf k+q'} \hat{a}_{n_2, \bf k}\0 (\hat{b}_{\bf q}\0 + \hat{b}^\dag_{- \bf q}) \,.
\end{eqnarray}
Here $\hat{b}_{\bf q}$ is the annihilation operator for the phonon, ${\bf q'}= (q_y,q_z)$ denotes the transverse component. $A$ is the matrix element for the transition between the harmonic oscillator energy levels:
\begin{equation}
 A(n_1,n_2,q_x) = \int \mathrm d x \, \varphi^*_{n_1}(x) \, \varphi_{n_2}\0(x) \, e^{- i q_x x} \, ,
\end{equation}
$\varphi_n(x)$ are the wave functions of the $n$-th eigenstate. Finally, the Hamiltonian of the phonons is  
\begin{equation}
H_{\mathrm{ph}} =  \int \frac{\mathrm d^3 {\bf q}}{(2 \pi)^3} \, \omega_q \, \hat{b}_{\bf q}^\dag \hat{b}_{\bf q}\0 \, , 
\end{equation}
where $\omega_q = c q [1 + (\xi q)^2/2]^{1/2}$ with the sound velocity $c$. Matrix elements for the impurity-phonon scattering are given by Eq.~(\ref{Vk}). 

In the present case the relative weakness of $a_{\mathrm{IB}}$ justifies a perturbative treatment of the problem. Our goal is the energy level renormalization for the impurities in the confinement potential. Its computation is best accomplished via lowest order irreducible self-energy correction.

 The energy level structure is given by the retarded self-energy, which is computed via analytical continuation of its Matsubara counterpart, see e.\,g.~ \cite{S_Mahan1990BOOK}, 

\begin{widetext}
\begin{eqnarray} \nonumber \nonumber 
 \Sigma(n, {\bf k}; i \Omega) = \frac{i \lambda^2}{\beta} \int \frac{\mathrm{d}^3 \bf q}{(2 \pi)^3} \sum_{i \epsilon_j}  \sqrt{\frac{(\xi q)^2}{ (\xi q)^2 + 2}}  \, \sum_{m=0}^\infty A^*(n,m,q_x)  A(m,n,q_x) \,\\
\times G[m,  {\bf k-q'}, i \Omega - (m-n) \om - i \epsilon_j] \, D_0(i\epsilon_j, {\bf q}) \, ,
\end{eqnarray}
where $\Omega$ is a fermionic/bosonic (impurity) and $\epsilon_j = 2 \pi j/\beta$ bosonic (phonons) Matsubara frequencies. $\beta=1/T$ is the inverse temperature. The necessary Green's functions are: 
\begin{eqnarray} 
 D_0(i\epsilon_j, {\bf q}) &=& \frac{2 \omega_{\bf q}}{(i \epsilon_j)^2 - \omega_{\bf q}^2} \\
 G[m,  {\bf k-q'}, i \Omega - (m-n) \om - i \epsilon_j] &=& \frac{1}{i \Omega - (m-n) \om - i \epsilon_j - E_{m,{\bf k - q'}}} \, ,
\end{eqnarray}
\end{widetext}
where the impurity Green's function $G$ has the same shape for both bosonic and fermionic case.

Performing the calculation in the assumption that only the two lowest lying energy levels are populated we can extract the energy difference between them. The result for the bosonic impurities is:
\begin{eqnarray}
\dSelf &=&  \frac{\Delta \omega}{\om} = \coup \cdot f(\rhoLiBEC, a, \xi)\,,\\
f(\rhoLiBEC,a, \xi) &=& \rhoLiBEC \xi^2 \nonumber \\ 
&&\times \left[ g_\mathrm{e}(a/\xi) \fexc -g_\mathrm{g}(a/\xi) (1-\fexc) \right]\,,\\
g_\mathrm{g}(a/\xi) &=& \frac{8 \pi^2}{\sqrt{2}} e^{(a/\xi)^2} \left[1-\erf(a/\xi) \right]\,,\\
g_\mathrm{e}(a/\xi) &=& 4 \pi \frac{a}{\xi} \sqrt{2 \pi} \nonumber \\
&&\times \left [1-\sqrt{\pi} \frac{a}{\xi} e^{(a/\xi)^2} (1-\erf(a/\xi)) \right]  \,.
\end{eqnarray}
The coupling strength $\coup$ is given by:
\begin{equation}
\coup = \frac{\lambda^2 m_\mathrm{B}}{8 \pi^2 \hbar^3 \om \xi}\, ,
\end{equation}
where $\hbar$ is the reduced Planck constant.

Self-energy evaluation for the case of fermionic impurities is slightly different. Here the numerical prefactors weakly depend on the effective temperature and the ratio $a/\xi$. Assuming $T=0$ and a symmetric superposition we obtain for the energy shift the value:
\begin{equation}
f(a, \xi) = -0.11+\frac{0.94}{a/\xi}\, .
\end{equation}
As the shaded inset in Fig.~2 left panel shows, the expression above is in very good agreement with our measurements.

\end{document}